\documentclass[letterpaper,twocolumn,10pt]{article}
\usepackage{usenix2020}
\usepackage{graphicx}
\usepackage{amsmath,amssymb}
\usepackage{algorithm}
\usepackage{algpseudocode}
\usepackage{booktabs}
\usepackage{xcolor}
\usepackage{booktabs}
\usepackage{tabularx}
\usepackage{makecell}
\usepackage{array}
\usepackage{url}
\usepackage{booktabs}
\usepackage{multirow}
\usepackage{graphicx}
\usepackage{hyperref}
\long\def\comment#1{}
\begin{document}

\title{\Large \bf LLM-Based Penetration Testing in the Presence of Honeypots}

\author{
{\rm Xinhong Xie, Piyush Nagasubramaniam, Neeraj Karamchandani, and Sencun Zhu}\\
{\small The Pennsylvania State University}\\
{\small \texttt{\{xjx5116,pvn5119,njk5270,sxz16\}@psu.edu}}
}

\date{}

\maketitle

\begin{abstract}

Large language model (LLM) agents are increasingly employed for offensive cybersecurity tasks such as automated vulnerability discovery, reconnaissance, and penetration testing.  This new capability also threatens one of the defender’s most valuable tools: deception. Traditional honeypots rely on realism and obscurity to lure human or script-driven attackers into revealing tactics, techniques, and procedures (TTPs), but LLM-driven attackers can reason about heterogeneous artifacts and use the honeypot suspicion to guide target-selection decisions.

We present a systematic study of honeypot-aware budget allocation for LLM attack agents. We formalize the attacker's problem as a budgeted decision process: an agent interacts with potential targets, consuming LLM execution budget during reconnaissance and
exploitation, and must decide whether to \emph{engage} (continue exploitation) or \emph{disengage} (skip) when honeypot suspicion arises.

Our findings show that with the proposed detector-guided policy, LLM agent attackers can effectively allocate budget to compromise hosts in a host pool, highlighting the importance of dynamically allocating budget in a controlled mixed-host testbed. While defenses are beyond our present scope, we discuss implications for future adversarially resilient and adaptive honeypot design.

\end{abstract}

\section{Introduction}
Deception has long been an important role in computer security. Honeypots and honeynets---controlled decoy systems designed to lure, engage, and study attackers---allow defenders to collect threat intelligence, observe adversarial tactics, techniques, and procedures (TTPs), and consume attacker resources. Traditionally, the effectiveness of such systems has depended heavily on their ability to appear sufficiently realistic to avoid detection and sustain attacker interaction.

LLM agents introduce a different attacker model. Unlike fixed penetration-testing scripts, an LLM agent can interpret reconnaissance results, select tools, revise intermediate decisions, and redirect its attack as new evidence becomes available. Unlike conventional scripts that may follow predetermined attack sequences, these agents can interpret observations, reassess intermediate results, and adjust their actions accordingly.

This adaptability changes the role of honeypot detection. A fixed attack sequence may continue interacting with a deceptive host once selected, whereas an LLM agent can use deception evidence to reconsider whether that host is worth additional attack cost. Because honeypots are designed to mimic legitimate targets and attract adversarial interaction, an attacker may spend a substantial portion of its limited execution budget probing or exploiting deceptive hosts rather than pursuing genuine targets. As the number of candidate hosts grows, exhaustively investigating every system becomes increasingly impractical. Effective autonomous attacks in such environments therefore require not only identifying exploitable systems, but also deciding \emph{which targets are worth further engagement under uncertainty and budget constraints}.

\subsection{Motivation}

Recent research on large language models has enabled increasingly autonomous agents for both defensive and offensive cybersecurity tasks. Defenders have incorporated LLMs into honeypots to generate more realistic interactions~\cite{Otal_2024,guan2024honeyllm} and to identify automated LLM attackers in the wild~\cite{reworr2025llmagenthoneypotmonitoring}. On the offensive side, systems such as PentestGPT and AutoPT demonstrate that LLM agents can reason across multi-stage penetration-testing workflows~\cite{299699,wu2024autoptfarend2endautomated}, while CHeaT investigates proactive defenses against such LLM-based attack agents~\cite{309754}.

These two trends expose an important but underexplored problem. Existing deception research primarily asks how defenders deploy honeypots, often modeling the attacker's ability to recognize deception as a fixed capability. Conversely, existing LLM-based offensive-security research largely focuses on vulnerability discovery and attack execution. Comparatively little is known about how an LLM attacker can actively infer whether a discovered target is deceptive and incorporate this uncertainty into its attack decisions.

Such capability is particularly important in resource-constrained attacks against environments containing both genuine and deceptive assets. Rather than treating every apparently vulnerable host equally, an attacker could use evidence gathered before and during interaction to estimate the likelihood that a target is a honeypot, prioritize higher-severity vulnerabilities targets, and avoid spending limited resources on deceptive branches. From the defender's perspective, this possibility also raises a fundamental question about how robust honeypot deployments remain when facing attackers that can actively reason about deception.

\subsection{Research Gap}

Existing work studies LLM-based penetration-testing agents, network-level deployment of deceptive resources, and honeypot detection and fingerprinting. However, it generally treats honeypot detection as an isolated classification or defensive problem. It does not examine how imperfect honeypot evidence should guide target selection and post-entry stopping when there is an LLM attacker operating under a finite execution budget.

Our work addresses this gap by integrating conservative pre-connect screening, budget-aware host ranking, and post-connect verification into a unified host-level attack policy. The novelty is not the detector itself. It is the decision-theoretic integration of imperfect honeypot evidence into sequential target selection under an LLM inference budget.

\subsection{Problem Statement}
We ask a fundamental question: 
\begin{quote}
\textit{Given a population of potential targets and a finite query budget,  how should an LLM-based attacker allocate its probing effort to maximize the number of confirmed genuine hosts while avoiding honeypots?}
\end{quote}

We formalize this as a constrained decision process balancing three factors: 
(1)~a normalized genuine-likeness score derived from honeypot evidence; 
(2)~a technical priority signal derived from vulnerability severity; and 
(3)~token-based cost of probing. 
Using local pre- and post-entry detection tools together with an LLM penetration-testing agent, we evaluate LLM agents’ ability to detect honeypots, measure token efficiency, and quantify resulting budget-allocation effectiveness.

\subsection{Contributions}
Our contributions are as follows:
\begin{itemize}
    \item We formulate honeypot-aware target selection as a
    budget-constrained decision problem for LLM attack agents.

    \item We propose a two-stage detector-guided policy that
    combines conservative pre-connect screening, budget-aware
    host ranking, and post-connect stopping.

    \item We evaluate the policy on a controlled pool of
    honeypots and genuine vulnerable hosts, decomposing its
    effects on detection, ranking, post-entry loss, and
    end-to-end budget efficiency.
\end{itemize}

\begin{figure*}[t]
\centering
\includegraphics[width=\textwidth]{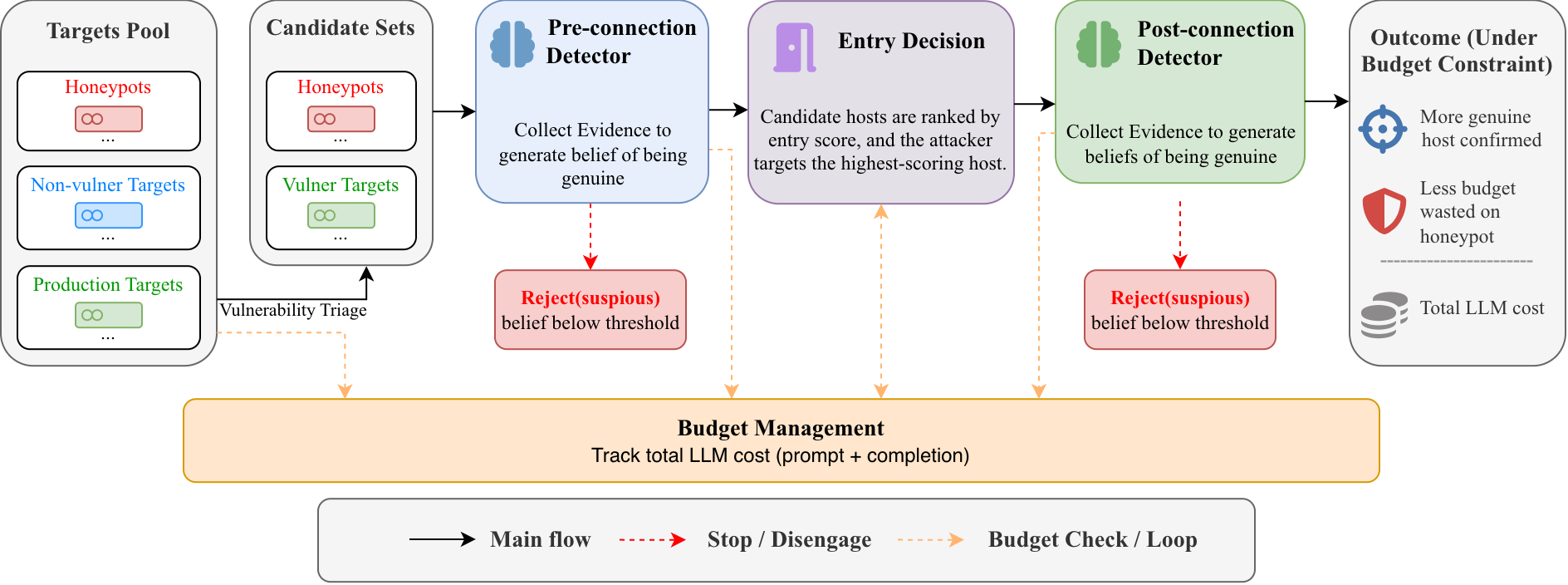}
\caption{Conceptual overview: an LLM agent performs budgeted reconnaissance across mixed honeypot and real targets, deciding whether to engage further or disengage based on artifact realism and cost.}
\label{fig:overview}
\end{figure*}

\section{Background and Related Work}
\label{sec:background_related}

\subsection{Honeypot Deception and Fingerprinting}

Honeypots are monitored decoy systems designed to attract attackers, collect adversarial behavior, and expose attack techniques~\cite{269618}. Prior surveys categorize honeypots by interaction level, deployment architecture, application
domain, and analysis goals~\cite{9520645,informatics12010014}. Its effectiveness depends in part on resistance to fingerprinting, because attackers may identify deceptive systems through banners, timing behavior, incomplete functionality, or other implementation-specific artifacts.

Recent work has applied LLMs to honeypot interaction and realism. LLM Honeypot uses the models to generate interactive web responses~\cite{Otal_2024}, while HoneyLLM supports LLM-generated shell interaction~\cite{guan2024honeyllm}. Other studies examine the defensive problem of observing LLM-based attackers in honeypot environments or systematizing potential honeypot-detection vectors~\cite{
reworr2025llmagenthoneypotmonitoring,
bridges2026sokhoneypotsllms
}.

Separately, honeypot-fingerprinting research has shown that implementation-specific protocol behavior can expose deceptive services. Vetterl and Clayton developed active probes for identifying low- and medium-interaction SSH, Telnet, and HTTP honeypots from their protocol responses~\cite{220564}. These results motivate the protocol-level pre-connect evidence used in our work.

\subsection{Threat Model}
\label{sec:threat_model}
We model an attacker faced with a huge population of targets $\mathcal{A} = \{a_1, \ldots, a_n\}$. After finishing a simple discovery/vulnerability triage, the attacker can get a host list $\mathcal{H} \subseteq A = \{h_1,...h_N\}$ that are vulnerable. Our study focuses on target selection and budget allocation within $\mathcal{H}$, rather than on the scalability of the upstream network-wide discovery process. Each host may be a vulnerable genuine system or a honeypot \(y_i \in \{0,1\}\). For each host $h_i$, the agent may perform different actions: collect honeypot evidence, or try to compromise the host, which costs $c$ tokens. After each action, the agent needs to decide what to do next and chooses among two options:
\begin{enumerate}
  \item \textbf{Continue engage/exploit} (incurring further cost and gaining more evidence, attempt compromise).
  \item \textbf{Disengage} (move on to another host).
\end{enumerate}
The attacker’s goal is to maximize expected compromised real machines under a total budget $B = \sum_t c_t \le B_{\max}$.  
Defenders are passive: they operate honeypots but do not hack back.  
We assume the attacker can access one LLM and use different tools for detection.

We consider defensively deployed honeypots whose attacker-controlled execution is subject to outbound containment. In particular, deceptive hosts are prevented from completing target-initiated connections to the controlled verification collector, whereas genuine hosts in our evaluation are configured to permit this connection.

This framework captures both economic and behavioral aspects of modern automated attacks and sets the stage for quantitative evaluation.

\section{Attacker Decision Model}
\label{sec:decision_model}

\subsection{System Overview}

As illustrated in Fig \ref{fig:overview}, the attacker operates over a heterogeneous target pool containing honeypots, non-vulnerable hosts, and genuinely vulnerable targets. A vulnerability triage step first identifies candidate hosts that are potentially exploitable. Before committing substantial attack resources, the pre-connection detector collects lightweight evidence to estimate whether each candidate is likely to be genuine; highly suspicious hosts are rejected without further engagement. For the remaining candidates, the LLM-based entry policy jointly considers the detector-derived genuine-likeness score, vulnerability severity, LLM cost already incurred on the host, and remaining budget to rank the hosts and select the next target to engage. After the attacker gains access to a selected host, the post-connection detector uses additional interaction evidence to reassess its authenticity and terminates further engagement if the host is suspected to be a honeypot. Throughout the process, the system continuously tracks LLM inference cost and repeats target selection and attack decisions until the available budget is exhausted. This design aims to maximize successful engagement with genuine targets while minimizing resources wasted on deceptive hosts under a fixed attack budget.

\subsection{Detector-Guided Host Decision}

Given the candidate host set $H$ defined in Section~ \ref{sec:threat_model},
the attacker maintains a normalized genuine-likeness score $b_i$ for each host $h_i$. The score is updated using pre-connect evidence before entry and post-connect evidence after a foothold is obtained. These signals determine whether a host is rejected, entered, or abandoned under the remaining LLM budget.

\subsection{Two-Stage Detection Decisions}

Before initial access, the pre-connect detector $D_{pre}$ maps externally observable protocol and service artifact evidence $E^{pre}$ to a normalized genuine-host score
\[
b_i^{\mathrm{pre}} = D_{\mathrm{pre}}(E_i^{\mathrm{pre}})
\in [0,1].
\]
Host \(h_i\) is rejected when the score is lower than the threshold $\theta^{pre}$
\[
b_i^{\mathrm{pre}} < \theta^\mathrm{pre},
\]
and otherwise remains eligible for budget-aware ranking.  Passing this stage does not confirm that a host is genuine, it only indicates that the available pre-connect evidence is insufficient for conservative rejection.

After a foothold is obtained, the post-connect detector $D^{post}$ uses execution and collector-observed execution evidence $E^{post}$ to produce
\[
b_i^{\mathrm{post}} = D_{\mathrm{post}}(E_i^{\mathrm{post}})
\in [0,1].
\]
The policy abandons the branch when the score is lower than the threshold $\theta^{post}$
\[
b_i^{\mathrm{post}} < \theta^{\mathrm{post}},
\]
and otherwise records or continues the branch as a genuine
target. The detector implementations are described in
Section \ref{sec:detector}.

\subsection{Budget-Aware Entry Scoring}

We account for all LLM inference cost incurred by the attack agent. At decision round $k$, the cumulative cost is
\begin{equation}
B_k = \sum_{j=1}^{k} \tilde{c}_j,
\end{equation}
and the episode terminates when $B_k \geq B_{\max}$.

The pre- and post-connect detectors, as well as other deterministic tools, execute locally and do not themselves incur LLM cost. However, they are exposed to the agent through the LLM tool interface. Therefore, model invocations used to select and issue a tool call and to process the returned tool observation are included in $\tilde{c}_j$ and charged to the episode budget.
Only the non-LLM runtime overhead of the underlying tool execution is excluded from $B_{\max}$.

After pre-connect screening, the attacker ranks the surviving candidates by an entry score. In our implementation, we use the normalized CVSS v3.1 Base Score \cite{first2019cvss31} as a heuristic proxy \(V_i\) for the technical severity of the vulnerability associated with host $h_i$. Let \(P_{hp} \geq 0\) be a conservative honeypot-entry penalty estimated from a held-out rule-based stopping baseline. Let \(0 \leq b^{pre}_{i} \leq 1 \) represent the normalized genuine-likeness score produced by the pre-connect detector.

With the above information, we define the risk-adjusted pre-entry utility score \textbf{utility} as
\[
    r_{i}
    =
    b_i^{pre}V_i
    -
    \left(1-b_i^{pre}\right)P_{hp}.
\]
The first term weights the vulnerability-severity score by the detector-derived genuine-likeness score, while the second term penalizes hosts exhibiting stronger honeypot evidence

The process of compromising a host still consumes LLM-agent execution budget. 
Let
\[
    \hat{c}^{spent}_{i,k} = \sum_{t\le k \\ a_t \to c_t }\tilde{c}_t,
\]
denote the cumulative LLM execution cost already incurred on host h
$i$ up to global decision round $k$. It contains only realized costs from previous actions on that host and therefore requires no knowledge of future attack cost.

We define the remaining budget-adjusted entry \textbf{score} as
\[
s_{i,k} = r_{i} - \mu_k \hat{c}^{spent}_{i,k}.
\]
Here, \(\mu_k \geq 0\) is a budget-pressure parameter at global decision round \(k\).

A larger \(\mu_k\) more strongly penalizes hosts on which more LLM execution budget has already been spent, encouraging the attacker to avoid repeatedly investing in costly targets under high budget pressure.

At each global decision round \(k\), the attacker selects the candidate with the highest positive entry score:
\[
    i^\star
    =
    \arg\max_{i \in h_i} s_{i,k},
\]

The action at round \(k+1\) will pick \(h_{i^\star}\) as the next target. Thus, the policy enters a host only when its budget-adjusted entry score remains the highest.

The score \(s_{i,k}\) has three roles. First, \(b^{pre}_{i,t}V_i\) prioritizes hosts that are both likely genuine and are associated with higher-severity vulnerabilities. Second, \((1-b^{pre}_{i})P_{hp}\) penalizes hosts whose evidence suggests honeypot risk. Third, \(\mu_k\hat{c}^{spent}_{i,k}\) penalizes hosts on which more LLM budget has already been spent, discouraging repeated investment in costly branches as budget pressure increases. The policy therefore does not merely classify hosts; instead, it ranks attack opportunities according to detector evidence, vulnerability severity, honeypot risk, and budget pressure.

\subsection{Budget-Pressure Update}

To adapt the ranking behavior under different budget regimes, we adopt the method from Bal. \cite{balseiro2023onlineresourceallocationhorizon} to update the budget-pressure parameter $\mu$ using a projected update rule:
\[
    \mu_{k+1}
    =
    \left[
        \mu_k
        +
        \eta
        \left(
            \tilde{c}_k - \lambda_k
        \right)
    \right]_+,
\]
where \(\eta>0\) is the step size, \([z]_+=\max(z,0)\), \(\tilde{c}_k\) is the realized LLM cost at decision round \(k\), and \(\lambda_k\) is the expected spending pace (expected cost at each round). When the attacker spends faster than the target pace, \(\mu_k\) increases and future entries are penalized more strongly. When spending is below the target pace, \(\mu_k\) decreases.

We define the expected spending pace as
\[
\lambda_k =
\begin{cases}
\dfrac{1}{1+\ln(\tau_2/\tau_1)}
\dfrac{B_{\max}}{\tau_1},
& k \leq \tau_1,\\[1.2em]
\dfrac{1}{1+\ln(\tau_2/\tau_1)}
\dfrac{B_{\max}}{k},
& \tau_1 < k \leq \tau_2,
\end{cases}
\]
where \(\tau_1\) and \(\tau_2\) denote the conservative and optimistic budget horizons, i.e., the estimated numbers of decision rounds supported under higher and lower per-round costs, respectively.

\section{Experimental Setup and Episode Construction}
\label{sec:experiment setup}

This section describes the controlled testbed and episode construction used to instantiate the budget-aware attack model in Section~\ref{sec:decision_model}. Unlike an artifact-level honeypot classification benchmark, our evaluation is organized around a population of candidate hosts. In each episode, the attacker is faced with a mixed host pool containing genuine vulnerable machines and honeypot deployments. The attacker must decide which hosts to enter under a limited LLM execution budget, using pre-connect evidence for conservative screening and post-connect evidence for continuation decisions after a foothold is obtained.

\subsection{Testbed and Target Pool}

We construct a controlled post-triage candidate pool containing 20 hosts: 8 honeypot deployments and 12 genuine vulnerable machines. This pool represents H in our threat model, i.e., the subset of hosts retained after inexpensive upstream discovery and vulnerability triage, rather than the complete asset population of a data center. All candidate hosts are deployed inside an isolated virtual network reachable only by the attacker environment. The target pool contains multiple service families, including SSH, Web, SMB, and FTP.

The honeypot subset includes Cowrie, Wetland, Glastopf, HoneySSH, sshForShits, sshsyrup, sshesame, and a second Cowrie configuration. These systems mainly emulate SSH, Telnet, and web-facing services and expose different levels of interaction fidelity.

The genuine subset consists of intentionally vulnerable laboratory machines from Vulnhub \cite{vulnhub}, including Metasploitable, DC-1, Kioptrix Level 1, Stapler, and other vulnerable virtual machines. A host is considered genuine when it represents a functional vulnerable system on which the attacker can obtain
verifiable progress, such as authenticated access, remote command execution, or an interactive shell.

\subsection{Detector Implementation}
\label{sec:detector}
\subsubsection{Pre-Connect Detector}
\label{sec:pre-connect-detector}

The pre-connect detector performs service scanning before starting real detection. After identifying the service protocol, the detector executes multi-stage probes. For each probe, it will collect implementation-level artifacts, including the server banner, packet lengths, padding patterns, responses to atypical packets, disconnect messages, and selected raw-byte features as a feature vector $x_{i,p}$ 

And the fingerprinting procedure is inspired by the active protocol-deviation methodology of Vetterl and Clayton \cite{220564}, which uses crafted handshake inputs and implementation-specific response features to distinguish honeypots from genuine services. 

Then, using the cosine similarities between $x_{i,p}$ and real host features $c^{hp} $ and $c^{gen}$, which are centroid features of the training set from different prototypes. To avoid data leakage, we used different hosts to construct the prototypes and evaluate the detector, but they may come from the same family.

$x_{i,p}$ and honeypot features to calculate the distance $d_{i,p}$, which shows whether the host is closer to the real host or the honeypot.

The relative similarity difference is

\[
d_{i,p}=\max \operatorname{cos}(x_{i,p},c^{hp}) - \max \operatorname{cos}(x_{i,p},c^{gen})
\]









The detector combines the two probe decisions conservatively:

\[
\hat{y}^{pre}_i =
\begin{cases}
\mathrm{honeypot},
&
d_{i,p}\geq 0
\\
\mathrm{genuine},
&
d_{i,p} < 0
\end{cases}
\]

In addition, it produces a continuous genuine-likeness score based on the weaker of the two probes:

\[
b^{pre}_i
= 1 - 
\sigma
\left(
\min
\left\{
d_{i,p}
\right\}
\right)
\]

where $\sigma(\cdot)$ is the sigmoid function. A score below $0.5$ indicates that both probes exceed their honeypot thresholds, while a score above $0.5$ indicates that at least one probe does not provide sufficient honeypot evidence.

The score $b^{pre}_i$ is a genuine-likeness score derived from protocol-level honeypot evidence, rather than a calibrated posterior probability or a confidence score for the predicted label. Honeypot-labeled hosts are rejected before LLM-guided entry, whereas genuine and uncertain hosts remain candidates for downstream ranking.

\subsubsection{Post-Connect Detector}

The post-connect detector is invoked only after the attacker obtains a shell-like foothold. It follows a two-stage cascaded procedure that combines a hard command-side validity gate with a continuous outbound-connectivity score.

In the first stage, the detector issues a predefined set of low-risk commands and examines whether the resulting shell behavior is consistent with a functional operating system. The checks include command availability, output consistency, filesystem and process information, state persistence, and plausible side effects across consecutive commands. Let \(g_i^{\mathrm{cmd}}\in\{0,1\}\) denote the result of this stage. If the responses are clearly templated, inconsistent, or stateless, we set \(g_i^{\mathrm{cmd}}=0\), assign the host a post-connect score of zero, and immediately abandon the branch.

Hosts that pass the command-side gate proceed to a controlled outbound-connectivity test. The target is instructed to send a sequence of payloads to a collector deployed inside the isolated evaluation network. Let \(n_i^{\mathrm{out}}\) denote the number of valid payloads received by the collector within the observation window. A larger value provides stronger evidence that attacker-issued commands produce externally observable effects permitted by the target's network environment. Under our containment model, such collector-reaching side effects are permitted for genuine hosts but blocked for honeypot deployments. We convert this count into a normalized outbound score

\[
s_i^{\mathrm{out}} = f(n_i^{\mathrm{out}}) \in [0,1],
\]

where \(f(\cdot)\) is a monotonically increasing normalization function. Larger values indicate stronger evidence that the shell supports genuine network execution rather than a shallow simulated response.

The final post-connect genuine-likeness score is

\[
b_i^{\mathrm{post}}
=
g_i^{\mathrm{cmd}} s_i^{\mathrm{out}}.
\]

Thus, any host that fails the first-stage validity checks receives \(b_i^{\mathrm{post}}=0\), whereas hosts that pass the first stage receive a score determined by the number of valid outbound payloads observed. The attacker marks the branch as genuine when

\[
b_i^{\mathrm{post}}
\geq
\theta_{\mathrm{cont}}^{\mathrm{post}},
\]

or marks it as a honeypot and abandons it otherwise.

\begin{table}[t]
\centering
\caption{Detector and policy parameters used in the
experiments.}
\label{tab:parameter-settings}
\begin{tabular}{lll}
\toprule
Parameter & Value & Selection method \\
\midrule
\multicolumn{3}{l}{\textit{Detector parameters}} \\
$k_{\mathrm{pre}}$
    & 2
    & Fixed probe cap \\
$\theta_{\mathrm{skip}}^{\mathrm{pre}}$
    & 0.50
    & Natural decision boundary \\
$\theta_{\mathrm{cont}}^{\mathrm{post}}$
    & 0.50
    & Heuristically fixed \\
\midrule
\multicolumn{3}{l}{\textit{Budget-policy parameters}} \\
$P_{\mathrm{hp}}$
    & 0.438
    & Calculated rule-stop loss \\
$\mu_0$
    & 10
    & Fixed initialization \\
$\eta$
    & 0.1
    & Sensitivity analysis \\
$\tau_1,\tau_2$
    & Budget-dependent
    & Expected decision range \\
\bottomrule
\end{tabular}
\end{table}

\subsection{LLM Attacker}

The attack component is implemented using an LLM-based penetration-testing agent. For each selected host, the agent receives the target address, discovered ports and services, available software version and vulnerability information, and the observations returned by previous tool executions. The agent is not given the ground-truth host type or the outputs used internally by the honeypot detectors.

Our attacker is based on the red team agent of CAI \cite{mayoral2025cai} which uses OpenAI o3-mini \cite{openai2025o3mini} to develop an attack agent (making a trade-off between capability and cost efficiency). 
Unless otherwise stated, the model is configured with a default temperature. The conversation history and tool observations are retained during the current attack attempt, allowing the agent to revise failed actions and adapt its subsequent decisions.

The agent interacts with the evaluation environment through a controlled tool interface. The available tools support network reconnaissance, service enumeration, command execution, exploit-script execution, session management, and, most importantly, the pre-connection detector and post-connection detector. During an attack attempt, the agent follows an iterative observe--reason--act process, following the general reasoning-and-acting interaction pattern introduced by ReAct \cite{yao2023reactsynergizingreasoningacting}. It first analyzes the current target information and previous tool outputs, selects the next reconnaissance or exploitation action, executes the action through the corresponding tool, and incorporates the resulting observation into its next decision.

We define an attack attempt as one bounded interaction of the LLM attacker with a selected host. An attempt begins when the agent receives the current host context and proceeds through an iterative observe--reason--act process.

Before a foothold is obtained, the agent performs reconnaissance and exploitation actions. Initial access is considered successful only when the attacker obtains a verifiable capability, such as an interactive shell, authenticated remote access, or remote command execution. Establishing a network connection, receiving a service banner, or creating a non-interactive session alone does not constitute a successful foothold.

Obtaining initial access does not immediately terminate the host interaction. Instead, a successful foothold transitions the branch to the post-connect verification phase. In the full policy, the agent invokes the post-connect detector through its tool interface and uses the returned evidence to determine whether the apparent foothold should be accepted as genuine or abandoned as deceptive. The host branch terminates after genuine confirmation, honeypot abandonment, a terminal execution failure, or exhaustion of the remaining episode budget.

For each model invocation, we record input and output tokens and convert them into monetary cost according to the model pricing configuration. All LLM invocations made during reasoning and tool use, including those used to invoke the pre- and post-connect detectors and process their returned observations, are charged to the primary episode budget. The runtime cost of the underlying local tools themselves is excluded.

\subsection{Parameter Settings}
\label{sec:parameter-settings}

Table~\ref{tab:parameter-settings} summarizes all detector and policy parameters. Unless otherwise stated, the same settings are used across all policies and budget regimes.

Let $\bar{c}_{\mathrm{cmd}}$ denote the reference cost of one LLM-generated attack command. We compute the expected decision-range parameters as
\[
\tau_1 = \frac{B_{\max}}{20\bar{c}_{\mathrm{cmd}}},
\qquad
\tau_2 = \frac{B_{\max}}{\bar{c}_{\mathrm{cmd}}}.
\]
The decision-range parameters $\tau_1$ and $\tau_2$ are computed separately for each value of $B_{\max}$.

\section{Experimental Evaluation}
\label{sec:evaluation_plan}

Our evaluation is designed to test the full decision structure of the detector-guided two-stage attack policy. Rather than evaluating honeypot detection as an isolated artifact-classification task, we evaluate whether the proposed policy can allocate a limited attack budget effectively across a pool of candidate hosts. 

Importantly, in the real world, attackers are faced with a combination of honeypots, non-vulnerable machines, and vulnerable machines. To avoid wasting budget on non-vulnerable machines, in the CVE scan step, we take the machines has no CVE results or poor results as non-vulnerable machines and skip them. Each CVE scan only takes 0.002\$ to 0.003\$, even if we have 1000 non-vulnerable machines in a real scenario, the budget only increases by 2\$ to 3\$

To simplify, we evaluate our method on a testbed with 20 hosts: eight honeypot deployments and twelve intentionally vulnerable machines. 

\subsection{Ground Truth and Outcome Definitions}

Each host has a ground-truth type $y_i$. A genuine host is counted as confirmed only when the LLM attacker obtains a verifiable foothold or another predefined success artifact. A honeypot is counted as falsely entered if it survives pre-connect screening and receives LLM attack budget.

We distinguish five episode outcomes. A true genuine confirmation occurs when a genuine host is entered and successfully verified. A false reject occurs when a genuine host is discarded by pre-connect screening. A false entry occurs when a honeypot survives screening and receives LLM attack budget. A delayed abandonment occurs when a honeypot is entered and consumes post-entry budget before being abandoned. An unresolved host is one that remains unentered because the budget is exhausted or because its entry score is non-positive.

These outcome definitions connect the experimental setup to the evaluation metrics in the next section, including confirmed genuine hosts under budget, wasted honeypot budget, honeypot entry rate, cost to first genuine confirmation, and budget efficiency.

\subsection{Research Questions}
\label{sec:research-questions}

\textbf{RQ1: Detector validity.}
Can the pre-connect and post-connect detectors provide reliable signals for distinguishing genuine hosts from honeypots?

\noindent\textbf{RQ2: Entry-ranking quality.}
Does the budget-aware entry score prioritize genuine hosts with higher-severity vulnerabilities earlier than simpler ranking strategies?

\noindent\textbf{RQ3: Post-entry loss reduction.}
Can the post-connect detector reduce the budget wasted after the attacker enters a honeypot?

\noindent\textbf{RQ4: End-to-end budget efficiency.}
Under the same LLM budget, does the full policy confirm more genuine hosts and spend less budget on honeypots?

\subsubsection{RQ1: Detector Effectiveness}

We first evaluate whether the detector outputs are sufficiently informative for their intended roles in the policy. Because the pre-connect detector is configured as a conservative screening mechanism, we focus on its honeypot recall and its false-rejection rate on genuine hosts. We additionally report accuracy and F1-score for comparison with simpler
pre-connect baselines.

These detector-level results are not the final objective, but they establish whether the beliefs are informative enough to support downstream policy decisions.

\paragraph{Pre-connect screening.}
Table \ref{tab:preconnect-performance} compares the proposed pre-connect detector with the banner-only (Detector only makes classification decision based on the host's banner information) and service-metadata (Detector makes decision based on service-metadata information, which includes service, version, port, os, and other information) baselines. Our detector improves accuracy from 0.70 to 0.80 and F1-score from 0.50 to 0.67 and recall from 0.38 to 0.50. More importantly, it reduces the false-rejection rate on genuine hosts from 0.08 to 0.00.

Compared to banner-only and service metadata solutions, our proposed method focuses on the protocol level. Since many low- and medium-interaction honeypots do not employ real protocols, protocol-level detectors can uncover compelling evidence to identify them as honeypots.

These results are consistent with the intended conservative role of pre-connect screening. The proposed detector does not attempt to reject every honeypot from externally observable evidence. Instead, it rejects only hosts for which both protocol probes provide sufficiently strong honeypot evidence. This design preserves all genuine hosts in the evaluated pool, at the cost of allowing several ambiguous honeypots to proceed to the entry-ranking stage. 

\begin{table}[htbp]
    \centering
    \caption{Pre-connection detection performance. HP Rec. denotes
    \label{tab:preconnect-performance}
    honeypot recall, while Gen. FRR denotes the false-rejection rate
    on genuine hosts.}
    \label{tab:pre_detection}
    \setlength{\tabcolsep}{3.2pt}
    \renewcommand{\arraystretch}{1.05}
    \scriptsize
    \resizebox{\columnwidth}{!}{%
    \begin{tabular}{lcccc}
        \toprule
        Method
        & Acc. $\uparrow$
        & F1 $\uparrow$
        & HP Rec. $\uparrow$
        & Gen. FRR $\downarrow$ \\
        \midrule
        Banner-only      & 0.70 & 0.50 & 0.38 & 0.08  \\
        Service metadata & 0.70 & 0.50 & 0.38 & 0.08 \\
        Our methods         & \textbf{0.80} & \textbf{0.67}
                         & \textbf{0.50} & \textbf{0.00} \\
        \bottomrule
    \end{tabular}%
    }
\end{table}

\paragraph{Complementary detector roles.}
Table~\ref{tab:detector-level-performance} compares the final pre-connect and post-connect detector outputs. The pre-connect detector obtains a precision of 1.00 and a recall of 0.50. This means that the hosts rejected by the detector are honeypots, but only a subset of all honeypots can be identified before entry. Its ROC--AUC of 0.81 further indicates that the continuous detector score remains informative for ranking hosts that are not directly rejected.

In the evaluated post-entry branches, genuine hosts were configured to permit communication with the verification collector, whereas honeypot deployments were subject to containment that prevented such communication. Under this controlled deployment condition, the post-connect detector was evaluated on 8 honeypot and 8 genuine post-entry branches and could separate all evaluated branches. The post-connect detector achieves substantially stronger performance, with an accuracy, precision, recall, and F1-score of 1.00 in the evaluated post-entry branches. This difference is expected because the post-connect detector observes shell behavior, command consistency, state persistence, and outbound execution evidence that is unavailable before a foothold is obtained. 

\begin{table}[htbp]
\centering
\small
\setlength{\tabcolsep}{5pt}
\renewcommand{\arraystretch}{1.15}
\caption{Detector-level performance on pre-connect evidence and post-connect evidence under the controlled containment configuration.}
\label{tab:detector-level-performance}
\label{tab:detector_validity}

\begin{tabular*}{0.92\columnwidth}{@{\extracolsep{\fill}}lccccc@{}}
\toprule
Detector 
& Acc. $\uparrow$
& Prec. $\uparrow$
& Rec. $\uparrow$
& F1 $\uparrow$
& AUC $\uparrow$ \\
\midrule
Pre-connect 
& 0.80
& 1.00
& 0.50
& 0.67
& 0.81
\\

Post-connect 
& 1.00
& 1.00
& 1.00
& 1.00
& 0.98 \\
\bottomrule
\end{tabular*}
\end{table}

\subsubsection{RQ2: Entry-Ranking Quality}

We next evaluate whether the proposed entry score
\[
s_{i,k}
=
b_i^{pre}V_i
-
(1-b_i^{pre})P_{hp}
-
\mu_k\hat{c}_{i,k}^{\mathrm{spent}}.
\]
produces a better host-entry ordering than simpler alternatives.

This section answers the question: \emph{among hosts that survive pre-connect screening, does the score rank the highest-priority candidates earlier?}

We compare the proposed score with random, CVSS-only, detector scores-only, and partial-score baselines using Top-k genuine count, entry precision, and false-entry rate.

\paragraph{Effect of pre-connect detector outputs.}
Table~\ref{tab:entry-detector-comparison} reports the entry quality obtained using detector scores from different pre-connect methods. The proposed detector places three genuine hosts among the top three candidates, compared with two for the no-detector and service-metadata settings and one for the banner-only baseline.

The proposed detector also achieves the highest entry precision of 0.75 and the lowest false-entry rate of 0.25. Without detector guidance, entry precision is only 0.60 and the false-entry rate is 0.40. These results show that the continuous detector output remains useful even when it is not strong enough to reject a host directly: it helps concentrate genuine hosts near the top of the candidate ranking.

\begin{table}[htbp]
    \centering
    \caption{Entry-decision quality using detector scores produced by different pre-connection methods. All methods use the same vulnerability-severity scores, honeypot penalty, and cost estimates.}
    \label{tab:entry-detector-comparison}
    \setlength{\tabcolsep}{2.5pt}
    \renewcommand{\arraystretch}{1.05}
    \scriptsize
    \resizebox{\columnwidth}{!}{%
    \begin{tabular}{lccccc}
        \toprule
        Method
        & Top-3 Gen. $\uparrow$
        & First Rank $\uparrow$
        & Entry Prec. $\uparrow$
        & False Entry $\downarrow$
         \\
        \midrule
        No detector      & 2 & 1 & 0.60 & 0.40 \\
        Banner-only      & 1 & 1 & 0.69 & 0.31  \\
        Service metadata & 2 & 1 & 0.60 & 0.40 \\
        Proposed         & \textbf{3} & \textbf{1} & \textbf{0.75}
                         & \textbf{0.25}  \\
        \bottomrule
    \end{tabular}%
    }
\end{table}

All methods prioritize the genuine host machine. Although our detector does not improve the ranking of the first genuine host, it excels at enhancing the quality of subsequent candidate hosts and reducing the proportion of honeypots among the selected hosts.

\paragraph{Entry-score ablation.}
Table~\ref{tab:ranking-ablation} evaluates the components of the proposed entry score. CVSS-only ordering performs worse than random ordering, placing an expected 1.77 genuine hosts in the top three and 3.00 in the top five. In contrast, strategies incorporating the pre-connect detector scores place a genuine host first and achieve 3.00 genuine hosts in the top three.

Detector score-based results indicate that the pre-connect detector scores is the dominant signal for early prioritization. The full score does not further improve the first three positions, but increases the expected number of genuine hosts in the top five from 4.50 to 4.83. The budget-aware terms therefore mainly refine later entry decisions after the highest-confidence hosts have already been processed.

\begin{table*}[htbp]
\centering
\small
\caption{Entry-ranking quality among hosts that survive pre-connect screening.}
\label{tab:ranking-ablation}
\begin{tabular}{lcccc}
\toprule
Ranking Strategy (Expectation)
& Mean \# Genuine in Top-1 $\uparrow$
& Mean \# Genuine in Top-3 $\uparrow$
& Mean \# Genuine in Top-5 $\uparrow$\\
\midrule
Random-order: 
& 0.75 / 1
& 2.25 / 3
& 3.75 / 5\\

CVSS-only: $V_i$ 
& 0.60 / 1
& 1.88 / 3
& 3.00 / 5\\

Detector scores-based: $b^{pre}_{i}$ 
& 1.00 / 1
& 3.00 / 3
& 4.50 / 5\\

Full score: $s_{i,t,k}$ 
& 1.00 / 1
& 3.00 / 3
& 4.83 / 5\\
\bottomrule
\end{tabular}
\end{table*}

\subsubsection{RQ3: Post-Connect Loss Reduction}

A key design variable in the proposed policy is the honeypot loss
\[
L_{\mathrm{hp}}
=
\mathbb{E}\!\left[
\sum_{\tau=t_i^{\mathrm{entry}}+1}^{T_i^{\mathrm{abandon}}}
c(a_{i,\tau})
\;\middle|\;
i\in\mathcal{H}
\right].
\]
This quantity captures the expected additional budget wasted after entering a honeypot.

Here, $c(a_{i,\tau})$ denotes the LLM inference cost associated with agent step $a_{i,\tau}$. For a tool-mediated step, this cost includes the model inference required to issue the tool call and to process the returned observation, but excludes the runtime cost of executing the local tool itself.

We therefore evaluate whether the post-connect stage actually reduces realized post-entry waste. We compare no stopping, rule-based stopping, and the proposed post-connect stopping mechanism using average honeypot cost and stop rate.

This layer directly connects the empirical evaluation to the modeled term \(L_{\mathrm{hp}}\), showing whether the continuation gate meaningfully limits wasted downstream interaction.

\paragraph{Stopping performance.}
As shown in Table~\ref{tab:postconnect-stopping}, continuing without a stopping mechanism incurs an average honeypot cost of 0.500. The rule-based baseline reduces this cost only slightly, to 0.438, and stops 50\% of the evaluated honeypot branches.

The proposed post-connect detector reduces the average honeypot cost to 0.042 and stops all evaluated honeypot branches. Relative to the no-stop policy, this corresponds to a cost reduction of
\[
\frac{0.500-0.042}{0.500}\times 100\% = 91.6\%.
\]
It also saves 0.458 monetary units per entered honeypot branch, compared with only 0.062 for the rule-based baseline.

\begin{table}[htbp]
\centering
\caption{Post-connect stopping quality after foothold.}
\label{tab:postconnect-stopping}
\footnotesize
\setlength{\tabcolsep}{3.5pt}
\renewcommand{\arraystretch}{1.0}
\begin{tabular}{lcccc}
\toprule
Policy & HP Cost$\downarrow$ & Aban. Cost$\downarrow$ & Saved$\uparrow$ & HP Stop Rate $\uparrow$ \\
\midrule
No stop   & 0.500 & N/A & N/A  & 0\% \\
Rule stop & 0.438 & 0.376   & 0.062 & 50\% \\
Post-stop & 0.042 & 0.042  & 0.458 & 100\%\\
\bottomrule
\end{tabular}
\end{table}

This result demonstrates the distinct role of the post-connect stage. Pre-connect screening attempts to prevent suspicious hosts from receiving LLM attack budget, but externally visible evidence cannot identify every honeypot. When an ambiguous honeypot survives screening and is entered, post-connect verification limits the downstream loss by terminating the branch shortly after shell-level evidence becomes available.

\subsubsection{RQ4: End-to-End Budget Efficiency}

Our primary evaluation is conducted at the episode level. We compare complete attack policies under fixed total budget \(B_{\max}\). We set \(B_{\max}\) to 1\$, 2\$, 3\$, 4\$ and 5\$ and ran three episodes on the five budget setups. In this experiment, we compare our method with random policy and fixbudget policy. For the random policy, the attacker will randomly choose one host as the target for every attack attempt, and for the fixbudget policy, the total budget is evenly allocated to each host, then the attacker will continue focusing on one host until running out of the allocated budget or confirming it's a honeypot or genuine host.
Our primary evaluation is conducted at the episode level. We compare complete attack policies under five fixed LLM budget levels,
\[
B_{\max}\in\{\$1,\$2,\$3,\$4,\$5\}.
\]
For each budget level, we run three episodes using the same experimental protocol and compare the proposed full policy against random selection and fixed-budget allocation.

\paragraph{Primary episode-level metrics.}
We compare policies using the number of confirmed genuine hosts, wasted honeypot budget, and honeypot entry rate under budgets of \$1, \$3, and \$5.
We report:
\begin{itemize}
    \item \textbf{Confirmed genuine hosts under budget:} number of genuine hosts correctly confirmed before budget exhaustion;
    \item \textbf{Budget to first genuine confirmation:} amount of budget spent before the first genuine host is successfully confirmed;
    \item \textbf{Wasted budget on honeypots:} total cost spent on hosts that are ultimately honeypots;
    \item \textbf{Budget efficiency:} number of confirmed genuine hosts per unit cost;
    \item \textbf{Post-screen honeypot entry rate:} the fraction of honeypots that receive LLM attack budget among those that survive pre-connect screening.
    \item \textbf{Cost-to-achieve(\(K\)):} minimum budget required to obtain \(K\) genuine confirmations.
\end{itemize}

\paragraph{Performance across budget levels.}
Table~\ref{tab:budget-regimes} shows how policy behavior changes as the available LLM budget increases from \$1 to \$5. At the lowest budget of \$1, all three policies confirm only $0.33$ genuine hosts on average. This suggests that when the available budget is sufficient for only a small number of LLM-guided attack actions, the cost of obtaining initial access dominates differences in target selection.

At \$2, the effect of detector guidance begins to appear, although it has not yet translated into a clear advantage in the number of confirmed genuine hosts. The full policy and random selection both confirm $1.33$ genuine hosts, compared with $1.00$ for fixed-budget allocation. However, the full policy obtains the lowest honeypot entry rate, $27\%$, compared with $41\%$ for random selection and $63\%$ for fixed-budget allocation. This indicates that under a still constrained budget, detector guidance first improves the quality of the hosts receiving attack budget before producing a substantial increase in total genuine confirmations.

The advantage becomes clearer at intermediate budgets. At \$3, the full policy confirms $2.33$ genuine hosts, compared with $1.66$ for both baselines, while also producing the lowest wasted honeypot budget ($0.16$). At \$4, the full policy further increases the number of confirmed genuine hosts to $2.66$, compared with $2.33$ for fixed-budget allocation and $1.66$ for random selection. Honeypot costs at this budget are similar across the three policies ($0.22$--$0.23$), indicating that the principal advantage of the full policy at this stage comes from allocating the available budget toward more productive genuine-host interactions rather than solely from reducing honeypot expenditure.

At the highest evaluated budget of \$5, the full policy confirms $3.66$ genuine hosts, compared with $3.00$ for fixed-budget allocation and $2.33$ for random selection. This corresponds to improvements of approximately $22\%$ and $57.1\%$, respectively. Because the instantiated attacker can obtain verifiable footholds on at most four genuine hosts in the current testbed, the full policy reaches approximately $91.5\%$ of this empirical capability ceiling at \$5. Thus, the five-point budget sweep shows a clear transition: policy differences are limited when the budget is extremely constrained, become increasingly consequential once multiple candidates can be processed, and eventually approach the capability ceiling of the underlying attacker.

The increase in honeypot entry rate at larger budgets should not be interpreted as a degradation of the detector. With more available budget, the attacker processes a larger fraction of the surviving candidate pool and therefore eventually reaches more ambiguous honeypots that were not rejected during pre-connect screening.

\begin{figure}[htbp]
    \centering
    \includegraphics[width=\columnwidth]{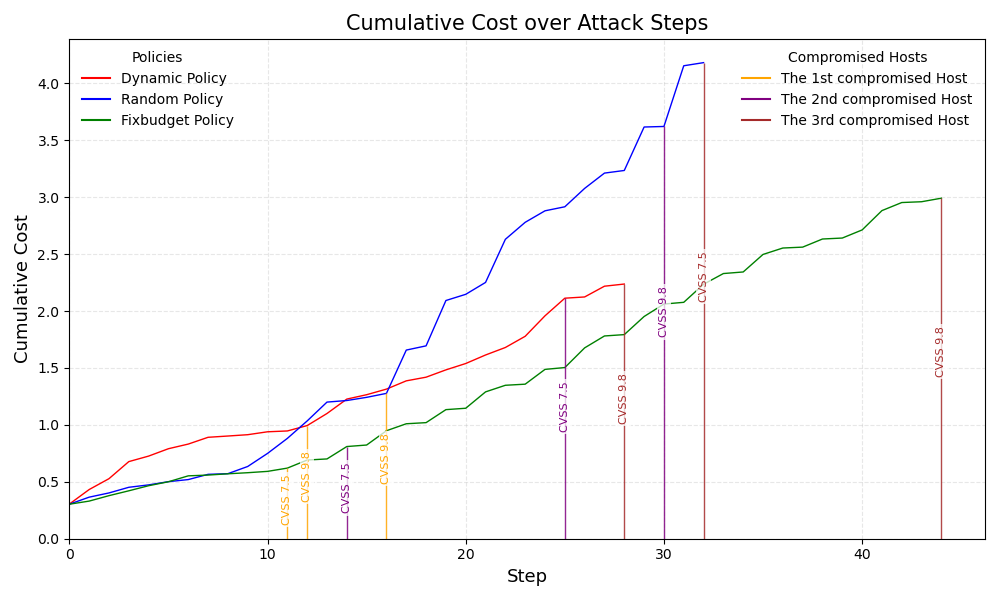}
    \caption{Cumulative LLM cost required to reach successive genuine-host confirmations. Vertical markers indicate the first, second, and third confirmed genuine hosts for each policy.}
    \label{fig:budget-efficiency}
\end{figure}

\paragraph{Cost to successive genuine confirmations.}
Figure~\ref{fig:budget-efficiency} complements the endpoint results in Table~\ref{tab:budget-regimes} by showing how much cumulative LLM budget each policy requires to reach successive genuine-host confirmations. For each policy, the vertical markers denote the attack steps at which the first, second, and third genuine hosts are confirmed; the corresponding value on the cumulative cost curve gives the cost-to-achieve that milestone.

The policies show relatively similar behavior for the first genuine confirmation, consistent with the ranking results in Table~5, where the main benefit of detector-guided allocation does not arise from the first target alone. Differences become more apparent for later confirmations. The detector-guided policy reaches the second and third genuine-host milestones with less cumulative LLM expenditure than policies that allocate attack effort without the same combination of deception evidence and dynamic budget pressure. This indicates that the principal benefit of the proposed allocation strategy is cumulative: after each host interaction, budget saved or reallocated from less productive branches remains available for reaching subsequent genuine targets.

\begin{table*}[htbp]
\centering
\small
\caption{Policy performance under different total budget regimes.}
\label{tab:budget-regimes}
\begin{tabular}{lccccc}
\toprule
$B_{max}$ 
& Policy 
& Confirmed Genuine / All $\uparrow$
& Budget to First Genuine (\$)$\downarrow$
& Wasted HP Budget $\downarrow$
& HP Entry Rate $\downarrow$\\
\midrule
1\$ 
& Random 
& 0.33 $\pm$ 0.47 / 12
& -
& 0.12 $\pm$ 0.06
& 38 $\pm$ 17\% \\

1\$
& Fixbudget 
& 0.33 $\pm$ 0.47 / 12 
& -
& 0.23 $\pm$ 0.01 
& 63 $\pm$ 0\% \\

1\$
& Full method 
& 0.33 $\pm$ 0.47 / 12 
& -
& 0.12 $\pm$ 0.01
& 33 $\pm$ 6 \% \\

\midrule
2\$ 
& Random 
& 1.33 $\pm$ 0.47 / 12
& 1.31 $\pm$ 0.33
& 0.13 $\pm$ 0.04
& 41 $\pm$ 11\% \\

2\$
& Fixbudget 
& 1 $\pm$ 0 / 12 
& 0.92 $\pm$ 0.24
& 0.24 $\pm$ 0.01 
& 63 $\pm$ 0\% \\

2\$
& Full method 
& 1.33 $\pm$ 0.47 / 12 
& 0.99 $\pm$ 0.35
& 0.14 $\pm$ 0.04
& 27 $\pm$ 3 \% \\

\midrule
3\$
& Random 
& 1.66 $\pm$ 1.24/ 12 
& 1.29 $\pm$ 0.50
& 0.25 $\pm$ 0.03 
& 58 $\pm$ 6\% \\

3\$
& Fixbudget 
& 1.66 $\pm$ 0.47 / 12 
& 0.90 $\pm$ 0.03
& 0.23 $\pm$ 0.02
& 63 $\pm$ 0\% \\

3\$
& Full method 
& 2.33 $\pm$ 0.47 / 12 
& 0.90 $\pm$ 0.05
& 0.16 $\pm$ 0.02  
& 41 $\pm$ 11\% \\

\midrule
4\$
& Random 
& 1.66 $\pm$ 1.24/ 12 
& 0.98 $\pm$ 0.38
& 0.22 $\pm$ 0.02
& 58 $\pm$ 6\% \\

4\$
& Fixbudget 
& 2.33 $\pm$ 1.24 / 12 
& 0.92 $\pm$ 0.03
& 0.23 $\pm$ 0.01
& 63 $\pm$ 0\% \\

4\$
& Full method 
& 2.66 $\pm$ 0.47 / 12 
& 0.96 $\pm$ 0.52
& 0.23 $\pm$ 0.05
& 58 $\pm$ 6\% \\

\midrule
5\$
& Random 
& 2.33 $\pm$ 0.94/ 12
& 1.34 $\pm$ 0.47
& 0.20 $\pm$ 0.02 
& 58 $\pm$ 6\% \\

5\$
& Fixbudget 
& 3 $\pm$ 0 / 12
& 0.89 $\pm$ 0.05
& 0.23 $\pm$ 0.02
& 63 $\pm$ 0\% \\

5\$
& Full method 
& 3.66 $\pm$ 0.47 / 12 
& 0.91 $\pm$ 0.03
& 0.18 $\pm$ 0.01
& 58 $\pm$ 6 \% \\

\midrule 

Upperbound & - & 4 /12 & - & -\\

\midrule

\end{tabular}
\end{table*}

\subsubsection{Ablation Test}

The preceding RQs evaluate different stages of the proposed policy separately and then assess their combined end-to-end effect. We further perform a leave-one-component-out ablation to determine whether each component contributes independently to the final policy behavior. We focus on the high-budget setting ($B_{max}$=\$5), because RQ4 shows that differences in target prioritization and continuation decisions become most visible when the attacker has sufficient budget to process multiple candidates. Starting from the full policy, we remove the pre-connect detector, the post-connect detector, or the dynamic budget-pressure mechanism while keeping the remaining components unchanged.

\paragraph{Ablation configurations.}
Starting from the full policy, we construct three leave-one-component-out variants while keeping the candidate pool, LLM attacker, host ordering, and total execution budget unchanged.

\textbf{Without pre-connect detector.}
We disable both the conservative pre-connect rejection gate and the detector-derived genuine-likeness score $b_i^{pre}$. Consequently, all post-triage candidate hosts remain eligible for entry. Because no host-specific deception evidence is available, the detector-dependent terms are removed from the entry score, yielding
\begin{equation}
s_{i,k}^{-\mathrm{pre}}
=
V_i-\mu_k\hat{c}_{i,k}^{spent}.
\end{equation}
The post-connect detector remains enabled. Thus, this variant isolates the contribution of pre-entry deception evidence while preserving vulnerability-severity prioritization, dynamic budget pressure, and post-entry verification.

\textbf{Without post-connect detector.}
We retain the pre-connect detector and the complete entry score, but disable post-connect authenticity verification. Once a verifiable foothold is obtained, no post-connect tool call is issued and the current host branch terminates without an additional honeypot-aware authenticity decision. Ground-truth host labels remain hidden from the policy and are used only by the evaluator to determine whether the apparent foothold corresponds to a genuine confirmation.

This variant therefore also avoids the LLM inference associated with invoking and processing the post-connect detector, while leaving the underlying episode budget unchanged.

\textbf{Without budget pressure.}
We retain both detection stages but remove the dynamic cost-pressure term by setting $\mu_k=0$ for all decision rounds. The entry score therefore reduces to
\begin{equation}
s_{i,k}^{-\mathrm{BP}}
=
b_i^{pre}V_i
-
(1-b_i^{pre})P_{hp}.
\end{equation}
The total episode budget $B_{\max}$ remains unchanged; this ablation removes only the mechanism that decreases a host's priority as LLM cost accumulates on that branch, rather than removing the budget constraint itself.

Table~\ref{tab:ablation} shows that removing each component degrades the full policy in a different way, revealing their complementary roles.

Removing the pre-connect detector slightly reduces the number of confirmed genuine hosts from $3.66$ to $3.33$, but substantially increases wasted honeypot cost from $0.18$ to $0.61$. The honeypot entry rate also increases from $58\%$ to $63\%$. Interestingly, the budget to the first genuine confirmation decreases from $0.91$ to $0.74$. This indicates that the pre-connect detector is not primarily responsible for reaching the first genuine host as quickly as possible. Instead, its main benefit is to improve the quality of candidates that receive downstream LLM budget by rejecting strongly suspicious hosts and providing a genuine-likeness signal for ranking. Without this evidence, the attacker may occasionally reach an early genuine target more quickly, but spends substantially more budget on deceptive hosts over the complete episode.

Removing the post-connect detector produces a different failure mode. The honeypot entry rate remains unchanged at $58\%$, which is expected because the post-connect detector operates only after a host has already been entered. However, the number of confirmed genuine hosts decreases from $3.66$ to $2.66$, the budget to the first genuine confirmation increases from $0.91$ to $1.28$, and wasted honeypot cost increases from $0.18$ to $0.31$. These results are consistent with the role of post-connect verification: it does not determine which hosts are entered, but provides additional evidence for distinguishing genuine progress from deceptive apparent footholds after entry. Removing this verification therefore leads to a less efficient downstream attack trajectory and leaves less effective budget for subsequent genuine targets.

The largest degradation occurs when dynamic budget pressure is removed. Setting $\mu_k=0$ reduces the number of confirmed genuine hosts from $3.66$ to only $0.33$. At the same time, this variant records no honeypot entries or honeypot-related cost. These zero values should not be interpreted as improved deception avoidance. Rather, without the cost-pressure term, accumulated expenditure on a host no longer lowers its priority, reducing the policy's ability to reallocate budget across candidate hosts. As a result, the attacker makes substantially less progress through the candidate pool before exhausting the available budget, which explains both the low genuine-confirmation count and the absence of observed honeypot entries.

Overall, the ablation results show that the three components address different aspects of budgeted target selection. The pre-connect detector improves candidate quality before entry, the post-connect detector verifies whether apparent progress after entry should be trusted, and dynamic budget pressure promotes reallocation away from costly branches. Their combination yields the highest number of confirmed genuine hosts while keeping honeypot-related expenditure substantially lower than the detector-ablated variants.

\begin{table}[t]
\centering
\caption{Component ablation under the high-budget setting ($B_{\max}=\$5$).}

\label{tab:ablation}
\resizebox{\columnwidth}{!}{
\begin{tabular}{lcccc}
\toprule
\textbf{Variant}
& \textbf{Confirmed Gen. $\uparrow$}
& \textbf{Budget to First $\downarrow$}
& \textbf{Wasted HP Cost $\downarrow$}
& \textbf{HP Entry Rate $\downarrow$} \\
\midrule

No pre-detector
& 3.33 $\pm$ 0.47 & 0.74 $\pm$ 0.07 & 0.61 $\pm$ 0.17 & 63 $\pm$ 0 \% \\

No post-detector
& 2.66 $\pm$ 0.47 & 1.28 $\pm$ 0.46 & 0.31 $\pm$ 0.07 & 58 $\pm$ 6 \% \\

No budget pressure
& 0.33 $\pm$ 0.47 & - & 0 $\pm$ 0 & 0 $\pm$ 0 \% \\

Full
& 3.66 $\pm$ 0.47 & 0.91 $\pm$ 0.03 & 0.18 $\pm$ 0.01 & 58 $\pm$ 6 \% \\
\bottomrule
\end{tabular}
}
\end{table}

\section{Discussion and Future Work}
\label{sec:discussion}

\subsection{Implications}

Honeypot effectiveness against LLM agents should not be evaluated only by whether a honeypot is eventually detected. A complementary measure is the amount of LLM execution cost imposed before the attacker reaches a confident decision. Our results show that pre-connect uncertainty and post-connect verification affect different stages of this cost.

\subsection{Ethical and Policy Considerations}

All experiments were conducted on authorized systems in an isolated laboratory network, and no external hosts were scanned or attacked. This study is limited to passive honeypot detection and budget allocation; active retaliation and interaction with third-party systems are outside its scope.

\subsection{Limitations}
Our evaluation is limited by the size and diversity of the controlled 20-host post-triage candidate pool, the concentration on SSH and web services, the small number of successful post-connect branches, and the exclusion of deterministic detector overhead from the primary LLM budget. The current study evaluates decision-making within a post-triage candidate set rather than end-to-end network-wide discovery or large-scale asset enumeration. Future work should evaluate larger and more heterogeneous candidate pools and integrate the policy with large-scale discovery or vulnerability-triage pipelines.

Future work should explore multi-agent co-evolution, online learning, and large-scale deployment traces.

\section{Conclusion}
\label{sec:conclusion}

We investigate budget allocation for an LLM attack agent operating over a mixed target pool containing both vulnerable genuine hosts and honeypots. We introduced a two-stage detector-guided policy that uses conservative pre-connect screening, budget-aware host ranking, and post-connect stopping.

In our testbed, conservative pre-connect screening preserved all genuine hosts while changing their priority relative to suspected honeypots. 
Within our controlled containment configuration, once a honeypot was entered, post-connect verification reduced the LLM budget spent continuing that branch. The end-to-end results further indicate that the benefit of the full policy depends on the available budget, highlighting the need to evaluate honeypot detection as part of an integrated attack process rather than as an isolated classification task.

Together, these results establish a host-level formulation of honeypot-aware budget allocation for LLM attack agents, while leaving larger and more heterogeneous data-center environments for future evaluation. Future work should evaluate larger target pools, more diverse services, and adaptive attackers and honeypots.

\bibliographystyle{plain}
\bibliography{references}

\end{document}